\documentclass[aps,prl,twocolumn,amsmath,showpacs,superscriptaddress]{revtex4}
\usepackage{graphicx}

\begin{document}        
 
\title{Wang-Landau algorithm for continuous models and joint density of states }
\author{Chenggang Zhou}
\affiliation{Computer Science and Mathematics Division, Oak Ridge National Laboratory, P.O. Box 2008, Oak Ridge Tennessee, 37831-6164 USA }
\affiliation{Center for Simulational Physics, University of Georgia, Athens Georgia, 30602 USA }
\author{T. C. Schulthess}
\affiliation{Computer Science and Mathematics Division, Oak Ridge National Laboratory, P.O. Box 2008, Oak Ridge Tennessee, 37831-6164 USA }
\author{Stefan Torbr\"ugge}
\affiliation{Universit\"at Osnabr\"uck, Fachbereich Physik, D-49076 Osnabr\"uck, Germany }
\author{D. P. Landau}
\affiliation{Center for Simulational Physics, University of Georgia, Athens Georgia, 30602 USA }
\date{\today}
 
\begin{abstract}

We present modified Wang-Landau algorithm for models 
with continuous degrees of freedom. We demonstrate this algorithm with
the calculation of the joint density of states $g(M,E)$ of ferromagnet 
Heisenberg models. 
The joint density of states contains more information than the density of 
states of a single variable--energy, but is also much more time-consuming 
to calculate. 
We discuss the strategies to perform this calculation efficiently for models 
with several thousand degrees of freedom, much larger than other continuous 
models studied previously with the Wang-Landau algorithm.

\end{abstract}

\maketitle

The Wang-Landau (WL) algorithm~\cite{Wang02} has been applied to a broad 
spectrum of interesting problems,
including statistical physics models such as Ising and 
Potts models~\cite{Wang02,Okabe}, spin glasses~\cite{Tomita}, and stochastic
series expansion of quantum systems~\cite{Troyer},
as well as models in other areas such as complex 
fluids~\cite{Shell} and protein molecules~\cite{Pablo}.
These applications have been successful due to two features of this
algorithm. 
First, the random walker of the WL algorithm is not 
trapped by local energy minima, which is very important for 
systems with complex energy landscapes. Secondly, by calculating the density
of states, thermodynamic observables in a wide range of temperature,
including the free energy, can be calculated with one single simulation. 
The efficiency of the WL algorithm has been quantitatively 
studied, and improvements have been proposed 
by several authors\cite{Shell, Dayal, Zhou03, Schulz03, Malakis},
which have made the WL algorithm a standard tool for discrete models. 
However, in some areas such as 
biophysics, most models have continuous degrees of freedom; and 
naturally, the density of states of two or more variables, in contrast to
the energy alone, is of much interest. Efficient algorithms for
those problems are called for.

In this letter, we discuss two generalizations of the WL algorithm: 
(1) physical models with continuous degrees of freedom 
(referred to as off-lattice models in Ref.~\cite{Shell}), and
(2) density of states of more than one 
variable~\cite{Shell,Zhou03, Landau04}, which we refer to as joint density 
of states. 
The joint density of states is more useful than the density of states. 
For example, the free energy as a 
function of both temperature and magnetic field can be calculated from
$g(M,E)$, the joint density of states of both energy and 
magnetization. For the protein system studied in Ref.~\cite{Pablo}, with
$g(\xi,E)$, where $\xi$ is the reaction coordinate, the free energy for each $\xi$ 
calculated with the EXEDOS algorithm can be obtained by integrating 
$g(\xi,E)e^{-E/T}$ over $E$. 
The calculation of joint density of states of continuous models clearly covers many
interesting problems.

\begin{figure}
\includegraphics[width=\columnwidth,height=0.45\columnwidth]{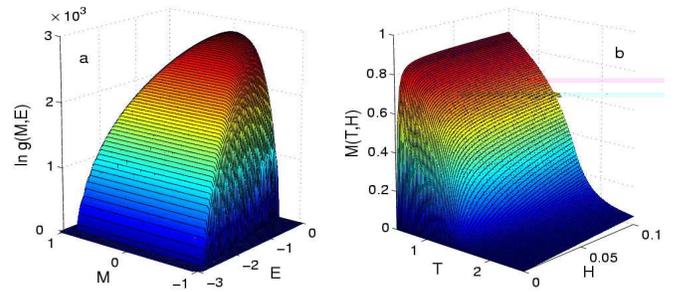}
\caption{(a)  $\ln g(M,E)$ of a three-dimensional Heisenberg ferromagnet 
of size $L = 10$ with cutoff energy at -2.8, 
determined upto an additive constant.
(b) The magnetization at different temperature and 
external field evaluated from $g(M,E)$.  }
\label{fig1}
\end{figure}

However, these generalizations are not straight-forward, and there are certain
intrinsic difficulties which frustrate the implementation of the simple 
version of the WL algorithm. First, the density of states $g(E)$ of a 
continuous model goes to zero at its maximum and minimum energies, which
become logarithmic singularities as the WL algorithm evaluates $\ln g(E)$.
In fact, $\ln g(M,E)$ is usually singular on its
boundary.
Secondly, suppose we calculate the joint density of a 32$\times$32 
Ising model on a square lattice, which was used as a test 
problem by Wang and Landau~\cite{Wang02}. If every value of the magnetization
is counted, the joint density of states $g(M,E)$ of both energy and 
magnetization will cost about $10^3$ times the CPU time of $g(E)$, 
because $g(M,E)$ contains about $10^6$ integers. Hence such 
calculations have been restricted to systems of small sizes~\cite{Landau04}. 
One might use a binning scheme to deal with this problem, and may 
also hope that binning (discretization) provides a generalization of the 
WL algorithm to systems with continuous degrees of freedom. However, 
a simple binning scheme becomes either very costly or inaccurate even for 
relatively small systems. Shell et. al.~\cite{Shell} used interpolation among
bins to handle this difficulty. We have identified the origin of the 
inefficiency of the binning scheme and propose a better solution to this 
problem.

We first show the result of our calculation for a three-dimensional Heisenberg 
ferromagnet in Fig.~\ref{fig1}. The Hamiltonian of this model is 
$H = -\sum_{<i,j>} {\bf S}_i\cdot{\bf S}_j$, 
where the summation goes over nearest neighbors on a cubic lattice of size 
$L$ with periodic boundary conditions. This model has a ferromagnetic phase 
transition and displays global spin rotational symmetry. Here we define 
$E = H/L^3$ and $M= L^{-3}\sum_iS^{z}_i$, 
and $g(M,E)$ is the joint density of states of $M$ and $E$. 
Figure~\ref{fig1}(a) shows $\ln g(M,E)$ for negative $E$. 
A region with $\partial \ln g / \partial M = 0$ below $E=-1.2$ indicates a 
transition to the ferromagnetic phase with a global rotational symmetry.
The logarithmic divergence in $w(M,E)$ near the ground state is obvious. 
In the following, we use the Heisenberg ferromagnet as a prototype model to 
identify several features and intrinsic difficulties of continuous models.

In general, the model we study has many microscopic degrees of 
freedom $s = (s_1,\ldots,s_N)$, which labels the microscopic states in the 
phase space $\Omega$. The Hamiltonian $H(s)$ is a real-valued function of the 
microscopic degrees of freedom, so is the order paramter, e.g. magnetization
$M(s) =  N^{-1} \sum_i S^z_i$ for the Heisenberg ferromagnet.
The joint density of states is defined as
\begin{equation}
g(M,E) = |\Omega|^{-1}\int\Pi_ids_i \delta(E-H(s))\delta(M-M(s)),
\label{eq1}
\end{equation}
where $|\Omega|$ is the volume of the phase space, and the integral is 
replaced by a summation for discrete models.
We refer to a pair $(M,E)$ as a macroscopic state, and define the macroscopic 
phase space 
$\Lambda = \{(M,E) |\, \exists s, H(s) = E, M(s) = M\}$. 
Obviously $g(M,E)\geq 0$ is a probability measure of $\Lambda$ induced by a 
priori probability measure of $\Omega$.
Usually this measure is almost a $\delta$-function in $\Lambda$, 
dominated by overwhelmingly many states in the high temperature limit. 
Due to the Boltzman distribution, for finite temperature properties,
we are mostly interested in the very small values of 
$g(M,E)$. In the following, we denote $(M,E)$ collectively with $x$. 
(In case of $g(E)$, the macroscopic state $x$ is specified by the 
energy alone.) 
If one tries to evaluate $g(x)$ by sampling $\Omega$ uniformly, then there
is virtually no chance to sample a finite temperature macroscopic state. 
The WL algorithm learns $g(x)$ in the runtime with a  
simple strategy. In a sentence, when a trial
move $x_i\rightarrow x_f$ is suggested, it is accepted with probability
$A_{i \rightarrow f} = \min\{1, \exp[w(x_i) - w(x_f)]\}$, where $w(x)$ is 
an approximation to $\ln g(x)$, and if the move is accepted $w(x_f)$ is 
updated with $\gamma + w(x_f)$ otherwise $w(x_i)$ is updated with 
$\gamma + w(x_i)$, where $\gamma = \ln f$, and $f$ is referred to as the
modification factor. Previous studies~\cite{Wang02, Zhou03} showed that 
for a discrete set of $x$, $w(x)$ quickly converges to $\ln g(x)$ within 
some statistical error proportional to $\sqrt{\gamma}$ 
for $\gamma \rightarrow 0$. For a continuous system, one simply cannot update
$g(x)$ point by point. Some modification of the update scheme is required
along with a method to represent the continuous function $w(x)$ in the computer
memory.

The simple binning scheme approximates $g(x)$ with a piece-wise constant 
function, which has discontinuities on the boundary of the bins. 
This scheme works if $g(x)$ varies very slowly in each bin, otherwise
as the trial moves within each bin is not biased,
the maximum point of $g(x)$ in each bin is much more likely to be sampled 
than other points, which actually violates the spirit of the WL algorithm--
to sample every macroscopic state uniformly.
Reducing the size of the bins could overcome this difficulty for small systems,
but generally for the joint density of states this would result in an 
excessively large number of bins to sample. 
Bilinear interpolation among neighboring bins was used by 
Shell et. al.~\cite{Shell} to prevent the random walker from being trapped 
in one bin. However, to capture the curvature of a 2D surface, a large number
of bins are still necessary. We will give an estimation of this number 
in later discussions.
Alternatively, we have adopted an update scheme similar to the kernel
density estimation~\cite{kde}(KDE), which is consistent with the continuous 
nature of $g(x)$. In our simulations of joint density of states, we select
a localized positive continuous kernel function $k(x)\geq 0$, 
and scale it by two constants $\gamma$ and $\delta$: 
$k(x) \rightarrow \gamma k(x/\delta)$.
If the random walker arrives in $x_0$, then the continuous histogram 
$w(x)$ is updated by
\begin{equation} 
w(x) \rightarrow w(x) + \gamma k( (x-x_0 )/\delta),
\label{eq2}
\end{equation} 
and we express the acceptance rate as 
\begin{equation}
A_{i\rightarrow f} = \min \biggl\{1, \exp\Bigl[\,\ln \alpha [\,w(x_i) - w(x_f)]\Bigr] \biggr\},
\label{eq3}
\end{equation} 
where we have inserted a constant $\ln \alpha$ so that $w(x)$ converges to 
$\log_\alpha g(x)$.
The continuity of $w(x)$ ensures that the random walker is always properly 
biased. This update scheme is similar to the method of Ref.~\cite{Wu04}, 
where the potential is updated by Gaussian kernels in a molecular dynamics 
simulation. 
In addition to the Gaussian kernel function $k(x) = \exp(-|x|^2)$, 
we also implemented the Epanechnikov kernel $ k(x) = [1-|x|^2]_+$. 
Little difference has been seen in the calculations with different kernel 
functions. By slightly modifying the approach in Ref.~\cite{Zhou03},
we can prove that this update scheme converges. 

Thus the single modification factor in the original 
WL algorithm is replaced by a triplet $(\alpha, \gamma, \delta)$. 
In our simulations, we have used numbers between 0.0001 and 0.01
for $\gamma$, and $\delta$ corresponding to about 1/200 of the width of the 
energy window or the magnetization window. Unlike the original WL algorithm,
we do not reduce $\gamma$ to extremely small values in the simulation, neither
do we change $\delta$ in the simulation. We have the freedom to change $\alpha$
provided that $w(x)$ is properly rescaled. A small $\alpha$ actually does not 
improve the accuracy of the calculation.

From Ref.~\cite{Zhou03}, we know if we start from an unbiased initial 
$w_0(x) = 0$, $w_T(x)$ (where $T$ labels the number of 
Monte Carlo steps) grows from the region of large $g(x)$, 
and extends to unexplored region of small $g(x)$. 
$w_T(x)$ can be written as:
$ w_T(x) = \bigl[\,C(T) + \log_\alpha g(x) + r_T(x)\bigr]_+$,
where $C(T)$ is an increasing function of $T$ only, $r_T(x)$ is a bounded 
error term controlled by the triplet $(\alpha,\gamma,\delta)$. 
$w_T(x)$ also increases monoticically in the simulation,
and remains zero in the unexplored region. One cannot expect it to be 
flat as in the original WL algorithm for discrete models. The simulation should
be stopped by other criteria, e.g. the visited area reaches a low energy 
cut-off. Then we continue the simulation with reduced $\gamma$ to improve the 
accuracy in the visited area. 
If the result is accurate, $w_T(x)$ increases uniformly in the 
visited area. $T$ is estimated by counting the number of 
kernel functions to build up $w_T(x)$:
\begin{equation}
  T = \left[\int \gamma k(x/\delta)dx \right]^{-1}\int_\Lambda w_T(x)dx .
\label{eq5}
\end{equation}

However we find that the initial accumulation in which 
$\Lambda_T = { \rm supp}\{ w_T(x)\}$ expands to $\Lambda$ takes a very long 
time for joint density of states of two variables. The expansion of $\Lambda_T$ 
slows down as the area of $\Lambda_T$ increases. The reasons are two-fold. 
First, the simulation samples the macroscopic states within $\Lambda_T$ 
uniformly, giving rise to a uniform growth there. This uniform growth takes a 
considerable CPU time, while only about a fraction $|\Lambda_T|^{-1/2}$ of 
Monte Carlo steps on the boundary of $\Lambda_T$ happen to extend the 
simulation to the unexplored area. Secondly, close to the singular boundaries 
of $\Lambda$, $\nabla \log_\alpha g(x)$ becomes very large,
requiring a very small $\delta$ 
(high resolution in the kernel function) 
to capture the large derivative. 

To avoid repeated sampling of the visited region  $\Lambda_T$,
and to push the simulation to the unexplored region, we find it is most efficient to 
introduce a global update of $w_T(x)$: when $w_T(x)$ is a good estimate of $\log_\alpha g(x)$ inside
$\Lambda_T$, we update $w_T(x)$ with the following formula:
\begin{equation}
w_T(x) \rightarrow w_T(x) + \kappa \exp\left[ { - \lambda\over w_T(x)-\omega }\right]\Theta(w_T(x)-\omega), 
\label{eq6}
\end{equation}
where $\Theta$ is the Heaviside step function.
Basically, $w_T(x)$ is shifted up by an amount of $\kappa$ where $w_T(x)>\omega$, and the 
exponential function removes the resultant discontinuity. 
Here $\omega$ is required to be positive because the above local update
scheme requires a minimum number of visits to give a correct estimation of the
density of states.

\begin{figure}
\includegraphics[width=0.85\columnwidth]{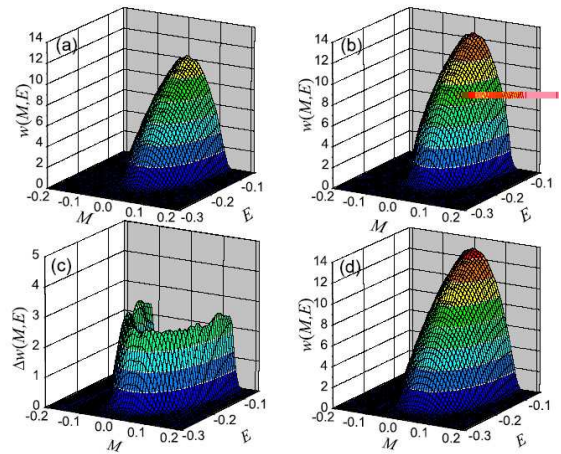}
\caption{A cycle of the simulation with global update. 
(a) original $w_T(x)$, 
(b) $w_T(x)$ after the global update, 
(c) the increment accumulated beginning with the $w_T(x)$ in (b), 
(d) sum of (b) and (c) gives a new $w_T(x)$ in the end of this cycle.}
\label{fig2}
\end{figure}

\begin{figure}[th]
\includegraphics[width=0.7\columnwidth]{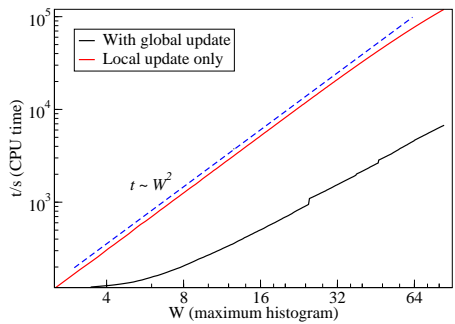}
\caption{Comparison of CPU time used by the calculations of $g(M,E)$ with and 
without the global update for an $L=5$ Heisenberg model of ferromagnet, $\omega = 0.5$ is used. 
The dashed line is a guide for the eye. }
\label{fig3}
\end{figure}

As a result of this global update, the random walker is confined close to 
$\partial \Lambda_T$, 
only when the accumulation on the boundary exceeds $\kappa$, is the random walker likely to
come back to the interior of $\Lambda_T$. Then a short global sampling using 
Eq.~(\ref{eq2}) covers up possible artifacts that result from the global update. 
Figure~\ref{fig2} illustrates one cycle of calculation with the global update. 
Ideally, the initial accumulation is decomposed into a 
number of cycles. In each cycle, the simulation works on a particular subset 
of $\Lambda$. 
Let $\Lambda_0 = \emptyset, \;\lambda_0 = \max\left\{\log_\alpha g(x) \right\}$,
$\Lambda_n = \left\{x | \log_\alpha g(x)>\lambda_0-n\kappa \right\}$ for $n>0$.
The series of sets $\{\Lambda_n\}$ converges to $\Lambda$. In $n$\textsuperscript{th} cycle 
of our calculation with the global update Eq.~(\ref{eq6}),
the random walker estimates $g(x)$ in $\Lambda_n-\Lambda_{n-1}$, 
which roughly requires 
\[ T_n = \left[\,\int \gamma k(x/\delta)dx \right]^{-1}
\int_{\Lambda_n-\Lambda_{n-1}} [\,\log_\alpha g(x) - \lambda_n] dx \]
Monte Carlo steps. By comparison with Eq.~(\ref{eq5}), we see that instead of filling
up the bulk of $w_T(x)$, the simulation only fills up a thin
surface layer of $w_T(x)$ of thickness roughly given by $\kappa$. 
Consequently, the algorithm with the global update saves a huge 
amount of Monte Carlo steps. Compared with distributing the random walkers to 
a number of ``windows''~\cite{Wang02,Schulz03, Shell},
the global update has the advantage of automatically selecting the ``windows'' on 
the frontier of the simulation and avoiding the boundary errors~\cite{Schulz03}.
Figure~\ref{fig3} shows that the global update saves 90\% of the CPU time in
a typical simulation, where we plot $t$ as a function of maximum histogram
$W \approx w(0,0)$. Without the global update, since the increment in $W$ is due to the uniform 
accumulation of samples inside the visited area $\Lambda_T$, therefore 
$dt/dW \propto |\Lambda_T|$. From Fig.~\ref{fig1}, we can tell that $\Lambda_T$ mainly grows 
towards lower energy. Hence we expect approximately $|\Lambda_T| \approx aW+b$, which results in a
leading quadratic dependence $t \propto W^2$ in Fig.~\ref{fig3}. With the global update, 
both the prefactor and the exponent of this relation are reduced. The short global samplings at 
the end of each cycle of the global update result in  $T\propto W^\lambda$, with 
$\lambda \approx 1.55$ in Fig.~\ref{fig3}.

We summarize the algorithm as the following: A calculation from scratch is 
divided into an initial accumulation stage and a refining stage. 
In the initial accumulation stage, 
({\bf 1}) we start the calculation with $w(x) = 0$, using local updates 
Eq.~(\ref{eq2}). 
({\bf 2}) As soon as  $w(x)>\omega$  for some $x$, we apply the global update 
Eq.~(\ref{eq6}), and continue the accumulation with local updates. 
$w_T(x)$ initially increases on the boundary of $\Lambda_T$. 
({\bf 3}) After it resumes a uniform growth over $\Lambda_T$, 
we start the next cycle with another global update.
The refining stage starts when $w_T(x)$ expands to the entire $\Lambda$ or 
certain cut-off one assumes. Then we reduce $\gamma$ (and $\delta$ if 
necessary) to continue the simulation with local updates only. 
When $w_T(x)$ resumes a uniform growth, we can further refine the result with 
reduced $\gamma$ or continue with the current parameters to decorrelate the 
statistical error and average over multiple uncorrelated results of $w_T(x)$.

\begin{figure}[thb]
\includegraphics[width=0.7\columnwidth]{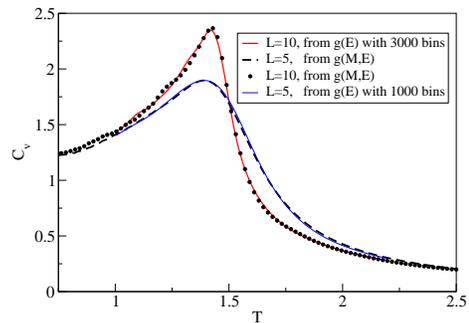}
\caption{ Specific heat of the Heisenberg ferromagnet of size $L=10$, and 5, with
comparison to the results from the original WL algorithm performed with a large
number of bins. }
\label{fig4}
\end{figure}

We calculate the thermodynamic quantities from $g(M,E)$ with numerical 
integral . Fig.~\ref{fig1}(b) shows the 
magnetization  of Heisenberg ferromagnet as a function of external field and temperature.
The specific heat in Fig.~\ref{fig4} shows a typical peak at $T_c$
of Heisenberg models. 
We also compare our results with that calculated with the 
original WL algorithm in Fig.~\ref{fig4}, 
which evaluates $g(E)$ on a grid of 3000 bins. They only differ
slightly at low temperatures where both results show small errors. 
This error comes from the binning or interpolation scheme used to represent 
the continuous $g(E)$ or $g(M,E)$ (not from the numerical integration). 
Actually, given that the standard deviation 
of the canonical distribution of the energy is $\sigma_E = \sqrt{T^2C_v/L^3} 
\approx 0.036$ for $L=10$, $T=1$, and $\sigma_E \approx 0.1$ for $L=5$, the 
numerical integration requires enough data points within $\sigma_E$ to be 
accurate. This criterion applies to both our kernel function updates and the 
bilinear interpolation scheme~\cite{Shell}.
A larger $\sigma_E$ explains why the error is very small for $L=5$ in 
Fig.~\ref{fig4}. In case of $L=10$, the internal array we used to store 
$g(M,E)$ has an energy resolution of 0.0012, which is comparable to the bin 
size (0.001) of the original WL algorithm we used for $g(E)$. 
Consequently, they show errors of comparable sizes. One
can conclude that for larger systems, the resolution in each macroscopic 
quantity is required to increase as $\sqrt{N}$ to maintain the accuracy in 
the numerical integral, where $N$ is the number of degrees of freedom. 

In summary, we have extended the WL algorithm to treate continuous systems 
and their joint density of states, proposed kernel 
function local updates and a global update to increase the efficiency of 
the algorithm. Our new strategies have potential applications
to many complex systems with thousands of degrees of freedom. In particular,
ther kernel function update benifits from the continuity of the model; 
and the global update effectively drives the random walker to unexplored area,
so that extreme values of macroscopic variables can be searched.  
If the calculations of $g(M,E)$ we presented were performed with the 
original WL algorithm, the CPU time used would have increased roughly by a factor of ten.

We thank D. Sanders for his helpful comments. 
This research is supported by the Department of Energy through the Laboratory Technology 
Research Program of OASCR and the Computational Materials Science Network of BES under 
Contract No. DE-AC05-00OR22725 with UT-Battelle LLC, and also by NSF DMR-0341874.

\end{document}